\documentclass[12pt]{article}
\usepackage{amsmath,amsfonts,amssymb,graphicx}

\newcommand\q[2]{\ensuremath{#1 \, \mathrm{#2}}}

\begin{document}
\begin{center}
  \textbf{\large The atomic structure of large-angle grain boundaries
  $\Sigma 5$ and $\Sigma 13$ in  ${\rm YBa_2Cu_3O_{7-\delta}}$  and their transport properties}\\[1cm]

    V.S.~Boyko, A.M.~Levine\\
  Department of Engineering Science and Physics\\
  The College of Staten Island, CUNY, Staten Island, New York 10314

\end{center}

\begin{abstract}
  We present the results of a computer simulation of the atomic
  structures of large-angle symmetrical tilt grain boundaries (GBs)
  $\Sigma 5$ (misorientation angles $\q{36.87}{^{\circ}}$ and
  $\q{53.13}{^{\circ}}$), $\Sigma 13$ (misorientation angles
  $\q{22.62}{^{\circ}}$ and $\q{67.38}{^{\circ}}$). The critical
  strain level $\varepsilon_{crit}$ criterion (phenomenological
  criterion) of Chisholm and Pennycook is applied to the computer
  simulation data to estimate the thickness of the nonsuperconducting
  layer ${\rm h_n}$ enveloping the grain boundaries.  The ${\rm h_n}$
  is estimated also by a bond-valence-sum analysis.  We propose that
  the phenomenological criterion is caused by the change of the bond
  lengths and valence of atoms in the GB structure on the atomic
  level. The macro- and micro- approaches become consistent if the
  $\varepsilon_{crit}$ is greater than in earlier papers. It is
  predicted that the symmetrical tilt GB $\Sigma5$ $\theta =
  \q{53.13}{^{\circ}}$ should demonstrate a largest critical current
  across the boundary.

  {PACS}: 74.50. +r; 74.72. -h; 61.72. Mm 

\end{abstract}

The major problem in the applications of ${\rm YBa_2Cu_3O_{7-\delta}}$
(YBCO) crystals is their low critical current density $J_c$ caused by
poor current transmission at the grain boundaries (GBs). The first
studies of this phenomenon, starting with
\cite{DimosChaudMannPhysRevLett88}, showed that $J_c$ across the GBs
decreases drastically with increasing misorientations angle $\theta$
between grains. Subsequent studies have indicated that certain
specific large-angle GBs (near-$\Sigma1$ $90^\circ$ [010] GBs) do not
show this sharp decrease (for details see
\cite{ChanGBJcY123JPhysChemSol95}).  Several mechanisms have been
suggested to explain the GBs influence on superconducting properties
of YBCO (see for details
\cite{GrossGBJJSpringer93,HilgencampMannhartIEETrans99}). We will
examine the explanations based on the reduction of the order parameter
caused by strain. Chisholm and Pennycook
\cite{ChisholmPennycookNature91} suggested that cores of dislocations
forming low-angle GB are nontransparent for the supercurrent because
superconductivity is suppressed in the region of the crystal lattice
where strains achieve some critical value $\varepsilon_{crit}$. Based
on this idea and using experimental data they derived the radius of
non-superconducting region surrounding dislocation $r^d_n$ as $r^d_n
\simeq 2.9 |\textbf{b}|$ (there $|\textbf{b}|$ is the magnitude of the
Burgers vector of the dislocation).  On this point the authors of
\cite{ChisholmPennycookNature91} left the semi-microscopical approach
and used a strain of $\q{1}{\%}$ as the cut off between YBCO being
superconducting and non-superconducting. This cut-off value was taken
from macroscopic consideration: from the fact that, as it is noticed
in \cite{BrowningBubanNellistNortonChisholmPennycookPhysicaC98}, a $1
\%$ strain causes YBCO to be tetragonal and so nonsuperconductive.
This phenomenological criterion allows one to estimate the thickness
of the non-superconducting layer enveloping the GB if the distribution
of the strain in the vicinity of it is known.  These type of estimates
were used for low-angle grain boundaries within the dislocation model
(see, for example, \cite{MeilikhovPhysicaC96}).
  
A precise physical meaning of the phenomenological criterion of
suppression of the superconductivity by strain is not well understood.
Within the approach \cite{GurevichPashitskiiPhysRevB98} the strain
shifts the chemical potential of the Josephson junction (JJ) barrier
at GB, and this may strongly affect the Josephson coupling, as long as
the barrier is characterized by the proximity to the metal-insulator
transition. Many properties of the low-angle GB JJ have been
successfully explained by this model
\cite{GurevichPashitskiiPhysRevB98}. It has been pointed out in
\cite{BrowningBubanNellistNortonChisholmPennycookPhysicaC98}, that
small changes of the local structure, existing in the vicinity of the
GB on the microscopic level, should induce large variations of the
valence of the copper ions, which control the transport properties.
This may give a plausible microscopic explanation for the exponential
drop in the critical current with grain boundary misorientation
\cite{BrowningBubanNellistNortonChisholmPennycookPhysicaC98}.
  
We propose that the phenomenological criterion of suppression of the
superconductivity by strain \cite{ChisholmPennycookNature91} is caused
by the change of the bond lengths, valence of atoms, and the number of
charge carriers present in the GB structure on the atomic level.
Therefore we will try to relate directly the bond-valence-sum analysis
to the phenomenological strain criterion of suppression of the
superconductivity. We use a very simple model and assume that all the
atomic bonds in the given point of the strained lattice experience the
same change of length irrespective of their orientation. Let us now
consider the strain of the interatomic bonds in the vicinity of the GB
and corresponding change of valence caused by it. Each bond between
atoms $i$ and $j$ is associated with the bond valence $s_{ij}$ that
can be determined accordingly to \cite{AltermattBrownActaCrystB85} by
the expression: $s_{ij} = \exp[(r_0 - r_{ij})/B]$, where $r_{ij}$ is
the bond length, $r_0$ and $B$ are empirical constants. The oxidation
state of atom $i$ (atom's valence) can be written as $V_i = \sum_j
s_{ij}$. If there is the average strain $\varepsilon$ in the region of
lattice in question, the bond length changes and can roughly be
estimated as $r^s_{ij} \approx r_{ij}(1 + \varepsilon)$. The atom
valence also changes to $V^s_i$.  It is easy to show that
$\varepsilon$, $V_i$, and $V^s_i$ are related by the following
approximate relationship:

\begin{equation}
  \label{eq:epsilon}
   \varepsilon \approx - \frac{B}{r_{ij}} \ln \frac{V^s_i}{V_i}
\end{equation} 

Now we can estimate $\varepsilon_{crit}$. If we take according to
\cite{BrowningBubanNellistNortonChisholmPennycookPhysicaC98} a valence
corresponding to the superconducting state as $V_i = 2.2$, and a
valence corresponding to the non-superconducting state as $V^s_i =
1.8$, according to \cite{AltermattBrownActaCrystB85} take $B = 0.37$,
and bond length between atoms Cu1 and O4 as $r_{ij} \approx 1.96 \AA$,
we will get the following approximate estimation $\varepsilon_{crit}
\approx 0.038$, that is, several times greater than the value of the
phenomenological criterion derived in \cite{ChisholmPennycookNature91}
from the macroscopical consideration.  Nevertheless these two
approaches become consistent if we will come back to the
semi-microscopical level of consideration. The strain field of
dislocation can be estimated accordingly to
\cite{KossevichTheCrystalLatticePhononsSolitonsDislocations1999} as
$\varepsilon \sim \frac{|\textbf{b}|}{4 \pi r}$ where $r$ is a
distance from geometrical center of dislocation. If we take $r = r^d_n
\simeq 2.9 |\textbf{b}|$, we would get $\varepsilon_{crit} \simeq
0.027$.  Thus we can deduce that the phenomenological criterion for
the suppression of superconductivity by strain at lattice level in
YBCO should be approximately $3\% - 4\%$.  The approach described
above can be used to estimate the nonsuperconducting region in the
vicinity of any type of the crystal lattice defects whose strain field
is known.  Therefore it is important to know the precise number of the
$\varepsilon_{crit}$.

  At the misorientations where the dislocation cores begin to overlap,
  the dislocation model of GBs becomes inapplicable (the concept of
  individual dislocations in the boundary becomes meaningless
  \cite{HirthLotheTheoryDislocations82}). Attempts to use the
  dislocation models to describe large-angle regime are not justified
  \cite{MeilikhovPhysicaC96}. Therefore there is no analytical
  method that allows a consistent calculation of the strain field for the
  large-angle GBs and then to estimate analytically the thickness of
  the nonsuperconducting layer at GB.
  
  In this work we apply the approach of Chisholm and Pennycook
  \cite{ChisholmPennycookNature91} for the most perfect large-angle
  GBs with smallest $\Sigma$ ($\Sigma$ is inverse of the density of
  coincident sites if lattices of neighboring grains are assumed to
  fill all space). Specifically, we get quantitative information about
  the atomic structure of these GBs by performing their computer
  simulation and determining the distribution of the strain in the
  vicinity of GBs. Then we use the phenomenological criterion to
  determine the thickness of the nonsuperconducting region enveloping
  large-angle GBs. The thickness of this region is estimated also by
  another method: determination of valences of atoms ${\rm Cu1}$ at
  the vicinity of GBs by analogy to
  \cite{BrowningBubanNellistNortonChisholmPennycookPhysicaC98}).
  
  An inter-atomic potential for ${\rm YBa_2Cu_3O_7}$ based on an
  unscreened rigid-ion model has been developed in
  \cite{ChaplotPhysRevB88}. This potential was used for computer
  simulation of twin boundary (TB) \cite{BoyGarKosRevPlast94} and TB
  and GB \cite{BoykoMalinskyAbdellatifBoykoPhysicsLettersA98} in ${\rm
    YBa_2Cu_3O_7}$. In \cite{ChaplotPhysRevB90}, the empirical
  interatomic potential model for ${\rm YBa_2Cu_3O_{7-\delta}}$ was
  proposed. We employ this potential \cite{ChaplotPhysRevB90} in this
  study as well as the Ewald method of calculation.
  
  We will characterize GBs by the direction of the axis of
  misorientation, the angle of misorientation relatively to this axis
  and the geometrical plane of GB which is chosen coinciding with some
  simple crystal lattice plane. In this article we present results of
  the computer simulation of the atomic structures of the symmetrical
  large-angle tilt GBs with the misorientation axes [001]: $\Sigma5$
  $(310)_1/(310)_2$, misorientation angle $\theta =
  \q{36.87}{^{\circ}}$; $\Sigma5$ $(210)_1/(210)_2$ $\theta =
  \q{53.13}{^{\circ}}$; $\Sigma13$ $(510)_1/(510)_2$ $\theta =
  \q{22.62}{^{\circ}}$; $\Sigma13$ $(320)_1/(320)_2$ $\theta =
  \q{67.38}{^{\circ}}$; $\Sigma29$ $(520)_1/(520)_2$ $\theta =
  \q{43.60}{^{\circ}}$. Boundary planes common for the two grains are
  given in parenthesis. The indices 1 and 2 refer to the two
  neighboring crystals of bicrystal. In general, the plane which is
  shared by the two crystals can be described by specifying the planes
  in each of the two crystals which is parallel with the common
  boundary plane \cite{TraeholtWenZandbergenShenHilgencampPhysicaC94}.
  The GBs $\Sigma 5$ and $\Sigma 13$ were chosen as the most perfect
  tilt GBs that cover wide range of misorientation angles and are
  experimentally observable
  \cite{BrowningChisholmPennycookNorthonLowndesPhysicaC93,TraeholtWenZandbergenShenHilgencampPhysicaC94}.
  The $\Sigma29$ GB is chosen as an example of GB close to the GBs of
  common type.
  
  The configurations of models are following. The ${\rm XOY}$-plane
  coincides with the basal ab-plane of the lattice, the ${\rm
    XOZ}$-plane coincides with the geometrical plane of GBs. The ${\rm
    X}$-axis lies at the GBs geometrical planes.  The ${\rm Y}$-axis
  is perpendicular to the GBs geometrical planes. The ${\rm Z}$-axis
  is directed along c-axis of crystal lattice and coincides with the
  misorientation axis. The extent of models along ${\rm X}$-axis in
  all cases are equal to one period of coincident-site-lattice of
  corresponding GBs. The total number of ions in model bicrystals for
  GBs $\Sigma5$ $\theta = \q{36.87}{^{\circ}}$, $\Sigma5$ $\theta =
  \q{53.13}{^{\circ}}$, $\Sigma13$ $\theta = \q{67.38}{^{\circ}}$ is
  624. For GBs $\Sigma13$ $\theta = \q{22.62}{^{\circ}}$, $\Sigma29$
  $\theta = \q{43.60}{^{\circ}}$ it is 650.
  
  Periodic boundary conditions were applied at all the outer faces of
  the computational cell. The molecular dynamics method was used. The
  equations of motion were solved using the velocity form of the
  Verlet algorithm with the starting time step not greater than
  $\q{2,5 \cdot 10^{-15}}{s}$. In each of cases of GBs simulation, we
  started with a sharp geometric initial configuration of the
  boundary.  It took us 224 time steps to achieve relaxed
  configuration in the case of GB $\Sigma 5$ $\theta =
  \q{36.87}{^{\circ}}$, 122 in the case of GB $\Sigma 5$ $\theta =
  \q{53.13}{^{\circ}}$, 2471 time steps in the case of GB $\Sigma 13$
  $\theta = \q{22.62}{^{\circ}}$, 4331 time steps in the case of
  $\Sigma 13$ $\theta = \q{67.38}{^{\circ}}$. In the case of GB
  $\Sigma 29$ $\theta = \q{43.60}{^{\circ}}$ the system was far from
  equilibrium after 3691 time steps and there are serious doubts that
  the equilibrium is achievable with a sharp geometric initial
  configuration of the $\Sigma 29$ GB. This means that, for such a
  boundary, the equilibrium could be reached only if a concentration
  of vacancies exists at the GB.
 
  Relaxed atomic structure of GB $\Sigma5$, $\theta =
  \q{53.13}{^{\circ}}$ is represented in the Fig. 1, GB $\Sigma13$,
  $\theta = \q{22.62}{^{\circ}}$ in the Fig. 2 (all distances are
  expressed in angstroms). The layer representation
  \cite{PooleFarachCreswikSuperconducitvity95} of the YBCO lattice
  cell is used. All layers are parallel to the basal plane. The plane
  consisted of Y atoms is a reflection plane.  Thus it is easy to
  imagine the atomic structure of rest lattice along Z-axis.
  Comparison of the initial and final (relaxed) configurations allowed
  to determine the atomic displacement vectors for the all atomic
  sites.  They are shown as arrows in the figures. It was found that
  for all GBs, the z-components of the atomic diplacement vectors are
  significantly smaller than x- and y-components. The maximal
  displacements are in the atomic plane next to the geometric plane of
  the GB. Especially large displacements have oxygen atoms of types O4
  and O3.  The magnitude of displacements drastically drops with
  distance from the GB geometric plane.

 \begin{figure}[t]
   \begin{center}
     \begin{tabular}{c@{\hspace{-1.75in}}c}
       \includegraphics[height=3.4in]{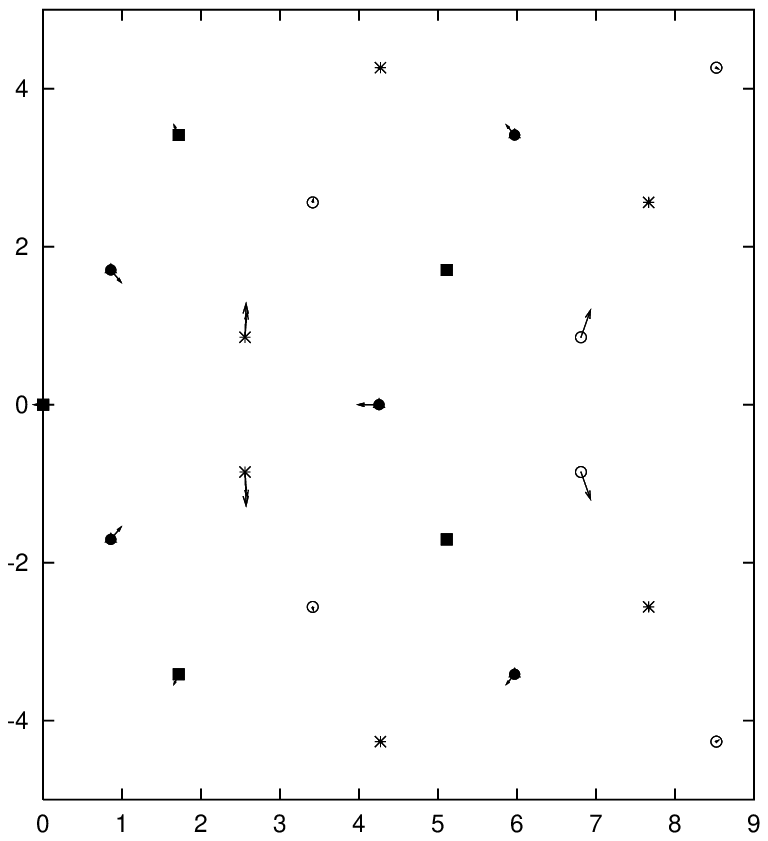} &
       \includegraphics[height=3.4in]{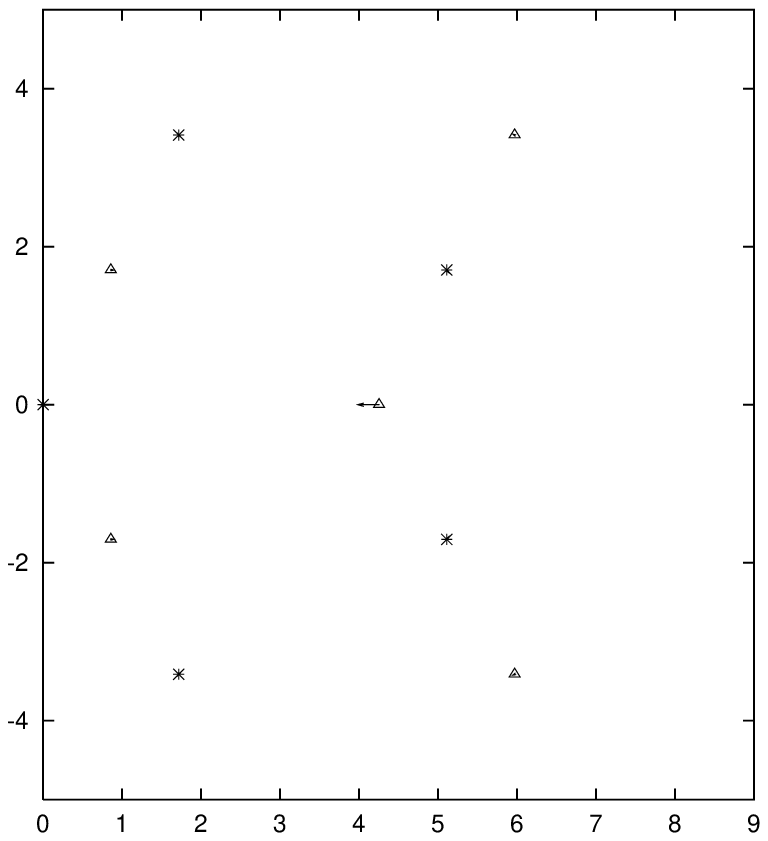}\\
       \hspace*{-1.5in} (a) & \hspace*{-1.5in} (b)\\[7mm]
       \includegraphics[height=3.4in]{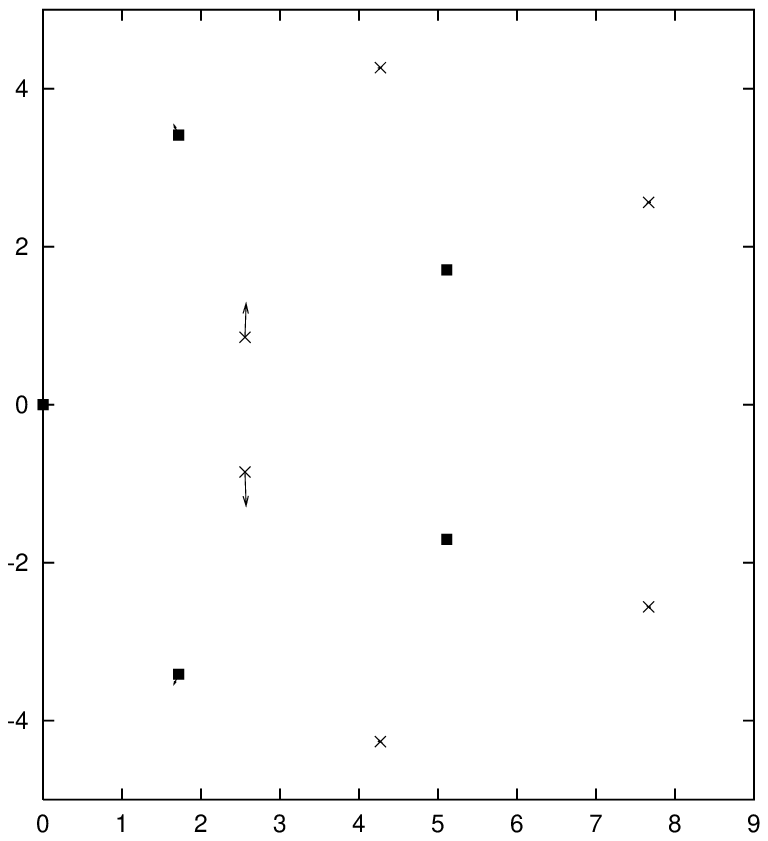} &
       \includegraphics[height=3.4in]{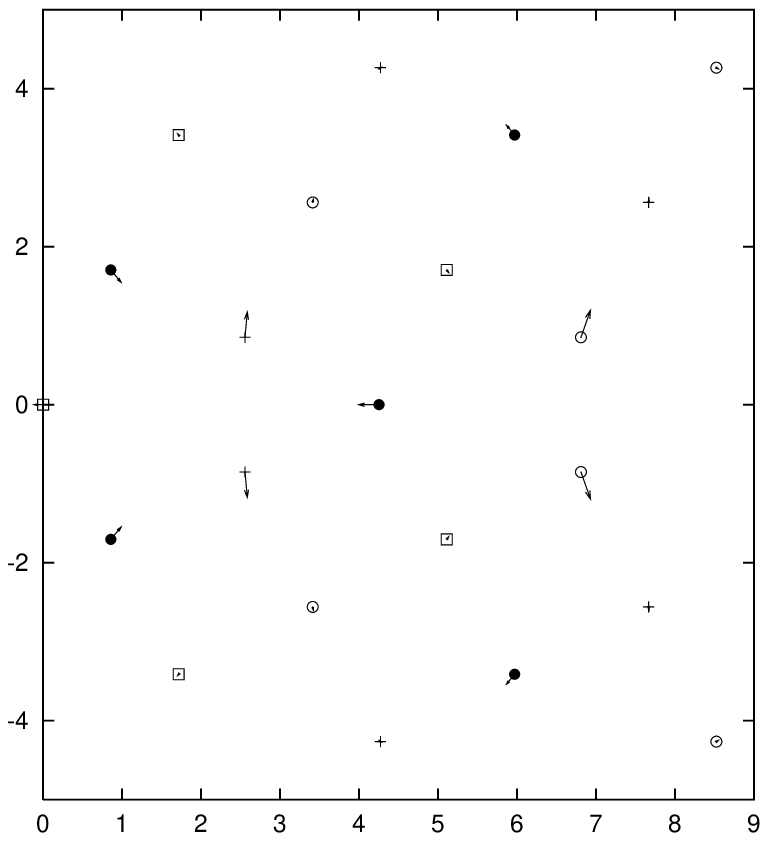}\\
       \hspace*{-1.5in} (c) & \hspace*{-1.5in} (d)
     \end{tabular}
     \caption[opi]{Atomic structure of GB $\Sigma 5$, $\theta =
       53.63^\circ$.\\ Y - +, Ba - $\times$, Cu1 - $\divideontimes$, Cu2
       - $\boxdot$, O1 -  $\blacksquare$, O2 - $\odot$, O3
       - $\bullet$, O4 - $ \triangle$.\\   The layer $\# 1$ (atoms Cu1
       and O4)is shown in Fig. 1(b). The layers $\# 2$ (atoms O1), $\#
       3$ (atoms Ba) are shown in Fig. 1(c). The layers $\# 4$ (atoms
       Cu2), $\# 5$ (atoms O2), $\# 6$ (atoms Y) are shown in Fig.
       1(d). The projections of layers $\# 1 - \# 6$ are shown in Fig 1(a).}
   \end{center}
 \end{figure}

 \begin{figure}[t]
   \begin{center}
     \begin{tabular}{c@{\hspace{0in}}c}
       \ \\[-0.8in]
       \includegraphics[height=3.5in,width=3in]{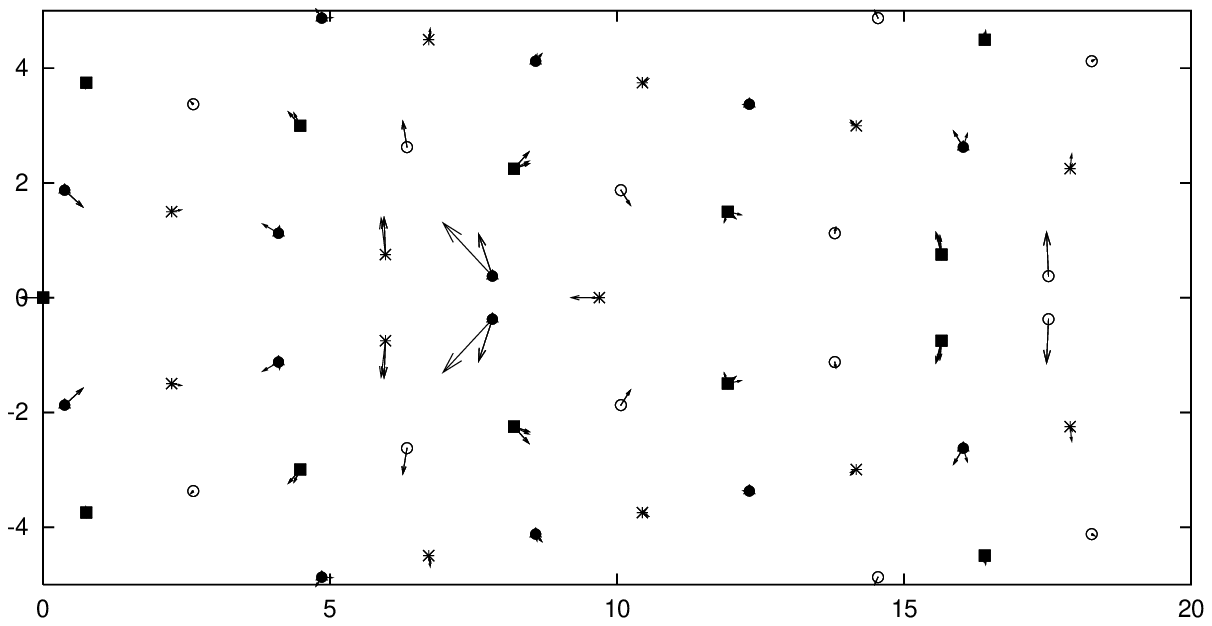} &
       \includegraphics[height=3.5in,width=3in]{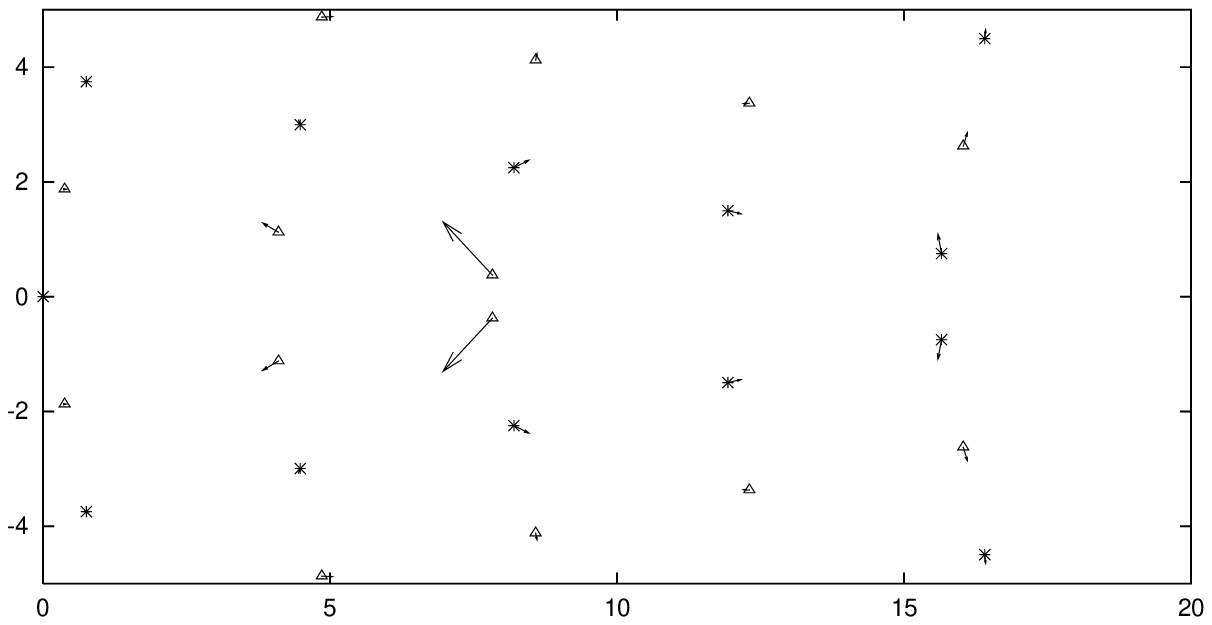}\\[-0.1in]
       (a) & (b)\\[-0.8in]
       \includegraphics[height=3.5in,width=3in]{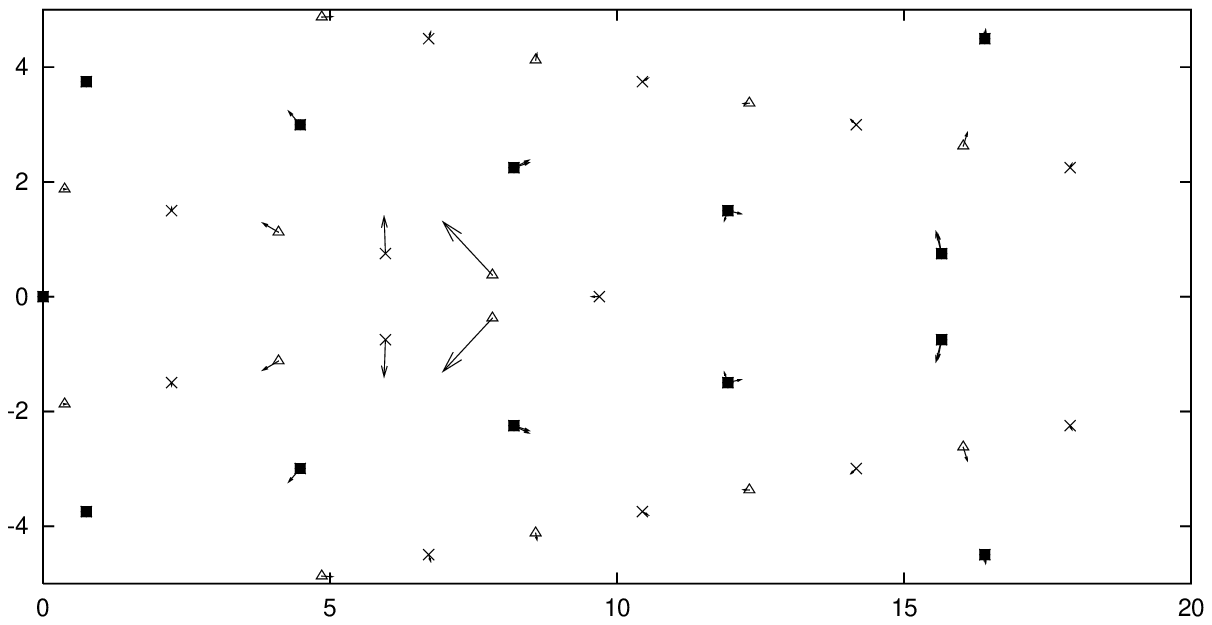} &
       \includegraphics[height=3.5in,width=3in]{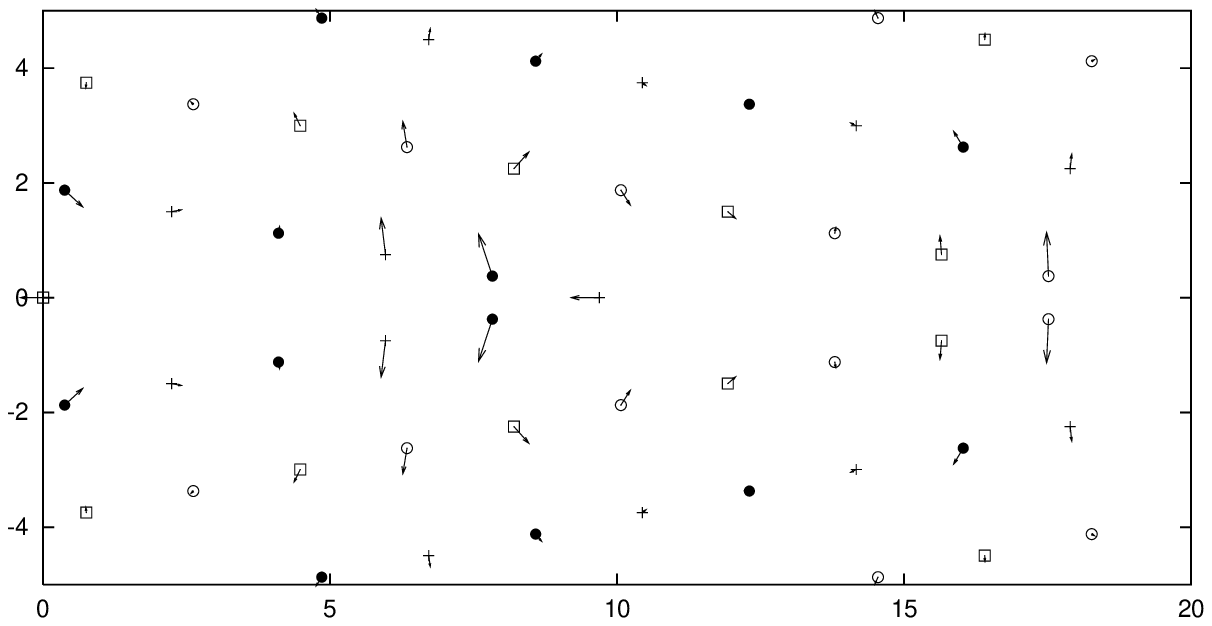}\\[-0.1in]
       (c) & (d)
     \end{tabular}
      \caption[opi]{Atomic structure of GB $\Sigma 13$, $\theta =
       22.62^\circ$.\\ The designations are the same as in Fig. 1.}

   \end{center}
 \end{figure}
 
 In order to facilitate analysis of spatial displacement distribution
 the average atomic displacement vector $ \textbf{u}$ was found in the
 each atomic plane parallel to the geometric plane of GB. The
 components of this vector were found by averaging of corresponding
 components of all the atoms at this atomic plane. For GB $\Sigma5$
 $\theta = \q{36.87}{^{\circ}}$, the magnitude of the average
 displacement vector $ u =| \textbf{u}|$ at the plane of the GB is
 $\q{1.54 \cdot 10^{-1}}{\AA}$. At the adjacent to the GB plane, $u =
 \q{5.53 \cdot 10^{-1}}{\AA}$, at the distance from GB $y =
 \q{4.23}{\AA}$ it drops by the order of magnitude, at the distance
 from GB $y = \q{6.64}{\AA}$ it drops by the two orders of magnitude
 and so on. The same character of displacement distribution is occured
 for GBs $\Sigma5$ $\theta = \q{53.13}{^{\circ}}$ and $\Sigma13$. The
 smallest displacements are in the vicinity of GB $\Sigma5$ $\theta
 =\q{53.13}{^{\circ}}$, the greatest in the vicinity of GB $\Sigma13$
 $\theta =\q{67.38}{^{\circ}}$. In the case of GB $\Sigma29$ the
 displacement of atoms ${\rm O3}$ and ${\rm O4}$ from sites adjacent
 to the geometrical plane of GB are extremely large $\approx
 \q{3}{\AA}$ and there is no evidence of their slowing down -
 configuration far from relaxed after 3691 time steps calculation. It
 seems that in relaxed configuration we would expect depletion of
 oxygen atoms at the GB.
 
 Our results agree with experimental observations of the atomic
 structure GB $\Sigma 5$ $\theta = \q{36.87}{^{\circ}}$
 \cite{BrowningChisholmPennycookNorthonLowndesPhysicaC93} that the
 symmetric $\Sigma 5$ $\theta = \q{36.87}{^{\circ}}$ GB demonstrates
 no detectable depletion zone and the lack of disordered region. The
 displacements of oxygen atoms outward of GBs and drastic increase of
 relaxation time for $\Sigma 29$ allows to suggest that the GBs with
 large $\Sigma$ could exist only if the oxygen depletion zone is
 formed in the vicinity of GBs. A presence of these zones, as it have
 been experimentally shown in
 \cite{WangKingZhuWagSuenagaPhilosMagA98,DuMironovaSathuamurthySalamaPhysicaC98},
 may essentially influence the superconducting properties of the
 large-angle GBs.
 
 Now we will try to relate atomic structure and transport properties
 of GBs estimating $h_n$ from computer simulation data.  The results
 of the bond-valence-sum analysis applied to the relaxed atomic
 structures of the GBs obtained by the computer simulation are
 represented in the Table. The distance from the GB plane where
 bond-valence sum becomes close to the same in the ``bulk'' is denoted
 as $r_n$ ($h_n$ = $2 r_n$). Then we will apply strain analysis
 \cite{BoykoMalinskyAbdellatifBoykoPhysicsLettersA98}.  The effect
 does not significantly depends on the type and sign of strain
 \cite{GrossGBJJSpringer93}. Therefore to facilitate analysis only the
 absolute value of the average atomic displacement $|\textbf{u}|$ in
 each atomic plane parallel to GB were used to construct approximate
 functions $|\textbf{u}(r)|$ ($r$ is the distance from the geometric
 plane of the GB).  A measure of the lattice strain $\varepsilon$ was
 estimated as the derivative of the approximate function with respect
 to $r$. The values of $h_n$ corresponding $\varepsilon_{crit} = 3 \%$
 are represented in the Table. Analysis of data shows for all cases
 that the smallest $h_n$ has GB $\Sigma5$ $\theta =
 \q{53.13}{^{\circ}}$. The GB $\Sigma5$ $\theta = \q{36.87}{^{\circ}}$
 has slightly greater $h_n$ though its misorientation angle is
 smaller. The $h_n$ of other considered GBs increases with the
 increasing of misorientation angles. The width of non-superconducting
 zone $ h_n $ can be used to estimate the expected tunneling current
 $J_c$ that exponentially decreases with $h_n$. One can expect that
 the more perfect special tilt GB $\Sigma5$ $\theta =
 \q{53.13}{^{\circ}}$ characterized by the smallest length of the
 periodicity along GB should demonstrate the larger $J_c$ across
 boundary. We can suppose that this GB has the best transport
 properties in comparison with other symmetrical tilt [001] GBs. The
 GB $\Sigma5$ $\theta = \q{36.87}{^{\circ}}$ should have smaller $J_c$
 though its misorientation angle is smaller. The two symmetrical GBs,
 which are complementary to each other, are different on an atomic
 scale and therefore, as it was pointed out in
 \cite{TraeholtWenZandbergenShenHilgencampPhysicaC94}, the electrical
 measurements need to be carried out on the whole range from
 $\q{0}{^{\circ}}$ to $\q{90}{^{\circ}}$. $J_c$ for the other analyzed
 GBs should decrease with the increasing of misorientation angle in
 agreement with the experiment. The results show that suggestion of
 suppression of superconductivity at GBs by strain leads to very sharp
 difference in $J_c$ through special GBs with small $\Sigma$. We
 suggest that this prediction could be tested experimentally (these
 results are relevant only for unfaceted GBs) and could shed light on
 applicability of the mechanism of suppression superconductivity by
 strain at the GBs in high-$T_c$ superconductors.
 
  {\em Acknowledgement:} We are grateful to A.B. Kuklov for useful discussions
  of our results.  This work is supported by CUNY Research Foundation
  Grant (Project $\# 62561-00-31$).

\newpage

\begin{table}[htbp]
  \begin{center}
    \leavevmode
    \begin{tabular}{|p{0.6cm}|p{1cm}|p{0.8cm}|p{1.3cm}|p{1.3cm}|}
         \hline
         $\Sigma$ & $\theta$ & GB Plane & $h_n ({\AA})$
         bond-valence-sum analysis & $h_n ({\AA})$
         strain analysis \\\hline
         $\Sigma13$ & $\q{22.62}{^{\circ}}$ & $(510)$ &
         $11.58$ & $9.88$ \\\hline
         $\Sigma 5$ & $\q{36.87}{^{\circ}}$ & $(310)$ & $9.48$ & $7.54$
\\\hline
$\Sigma 5$ & $\q{53.13}{^{\circ}}$ & $(210)$ & $7.06$ & $5.70$ \\\hline
       $\Sigma 29$ & $\q{43.60}{^{\circ}}$ & $(520)$ & $13.00$ &
        $10.56 $ \\\hline
       $ \Sigma 13 $ & $\q{67.38}{^{\circ}}$ & $(320)$ & $14.42$ &
       $11.46$ \\\hline
     \end{tabular}
   \end{center}
   \caption{Thicknesses of nonsuperconducting layers enveloping GBs
         with misorientation axis [001] in ${\rm YBa_2Cu_3O_{7-\delta}}$.}
   \label{tab:twin}
 \end{table}

\end{document}